\documentclass[10pt,twocolumn]{article}
\usepackage{times}

\usepackage{graphicx}
\usepackage{cite}

\usepackage{subcaption}
\usepackage{float}
\usepackage{xspace}
\usepackage{hhline}
\usepackage{multirow}
\usepackage{xcolor}
\usepackage{url}
\usepackage{hyperref}

\usepackage{makecell}%To keep spacing of text in tables
\setcellgapes{4pt}%parameter for the spacing

\usepackage[labelfont=bf]{caption}
\usepackage{hyphenat}

\newcommand{\cn}{\textsc{Sapphire}\xspace}
\newcommand{\TODO}[1]{}
\newcommand{\etal}{\textit{et al.}}

\begin{document}

\title{\cn: Automatic Configuration Recommendation for Distributed Storage Systems}
\author{Wenhao Lyu$^1$, Youyou Lu$^1$, Jiwu Shu$^1$, Wei Zhao$^2$ \\
\small {\em  $^1$Tsinghua University, \quad
          $^2$SenseTime Research \quad
          } \\ [2mm]
% \small Submission Type: Research
}
\date{}
\maketitle

\begin{abstract}
Modern distributed storage systems come with a plethora of configurable parameters that control module behavior and affect system performance.
Default settings provided by developers are often sub-optimal for specific user cases.
Tuning parameters can provide significant performance gains but is a difficult task requiring profound experience and expertise, due to the immense number of configurable parameters, complex inner dependencies and non-linear system behaviors.

To overcome these difficulties, we propose an automatic simulation-based approach, \cn, to recommend optimal configurations by leveraging machine learning and black-box optimization techniques.
We evaluate \cn on Ceph.
%We implement this system on Ceph distributed storage platform and provide general preprocessing strategies for such ambiguously defined parameters spaces.
Results show that \cn significantly boosts Ceph performance to 2.2$\times$ compared to the default configuration.

\end{abstract}
\section{Introduction}
\begin{figure}[!t]
    \centering
    \includegraphics[width=0.8\linewidth]{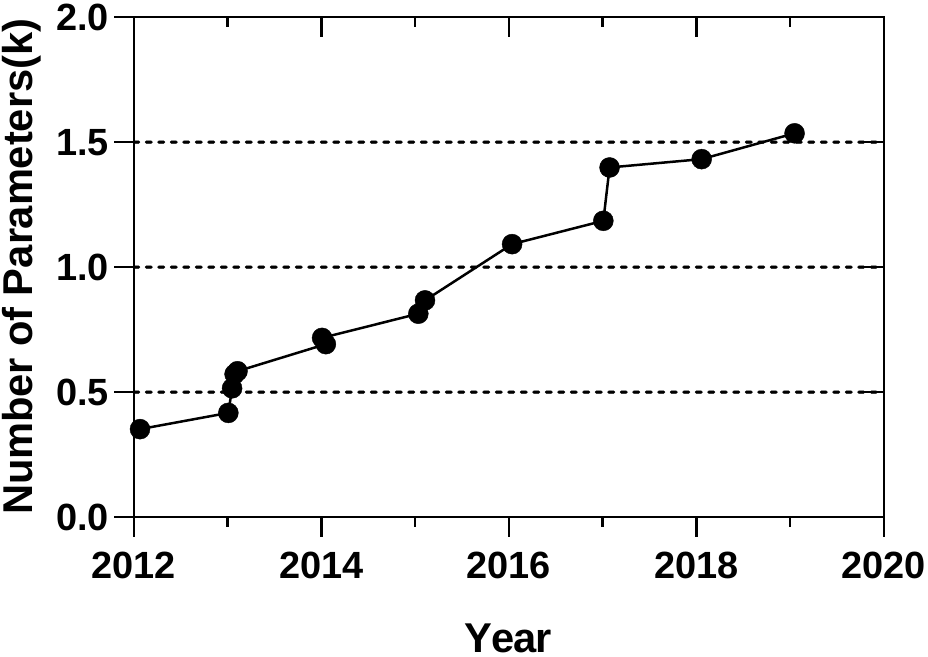}
	\caption{Ceph parameters growth. This figure demonstrate the number of parameters of major Ceph releases since 2012.}
    \label{fig:num_of_parameters}
\end{figure}

Modern distributed storage systems often have multiple layered and highly modular software architectures, support various types of user cases and consist of heterogeneous storage devices.
Construction of such a complex system comes with many design choices, producing a large number of configurable parameters~\cite{Tianyin2015}.
Figure~\ref{fig:num_of_parameters} depicts the number of Ceph parameters grows drastically to over 1500, near three times larger than the original version.
But storage systems are often deployed with default settings provided by developers,
rendering to be sub-optimal for specific user cases~\cite{Sehgal2010}.

Tuning parameters can provide significant performance gains~\cite{Zadok2015}, but is a challenging task for ordinary users, as rich experience and deep insight into system internals are required.
The performance impact of parameter settings is highly related to the hardware and workload characteristics.
There is no single configuration that can work well under all kinds of user cases~\cite{cao2018atc}. 
%be superior to all the others
However, depending on expert system administrators to manually tune parameter values for different user cases is unrealistic.
An automatic solution to generate near-optimal configurations for different user scenarios is in demand.

We find new challenges in distributed storage scenarios compared to previous studies on automatic parameter tuning in the storage system~\cite{cao2018atc,Zadok2015,cao2019inter}. 
% Compared to previous studies on automatic parameter tuning in storage systems~\cite{cao2018atc,Zadok2015,cao2019inter}, we find new challenges in distributed storage scenarios.
(1) \textbf{Configuration constraints.} 
%Ambiguous parameter space with hidden value constraints.
There exist many value constraints inside the parameter domain. Misconfigurations that violate some constraints can cause system failures or even crushes.
(2) \textbf{Huge numbers of parameters.}
Distributed storage systems often provide immense numbers of parameters. The newest Ceph Nautilus even comes with 1536 parameters.
Search for optimal settings in such enormous knob space is still challenging for popular black-box techniques.
%(3) Evaluations of distributed systems are very time consuming as restart or redeploy clusters are needed to make new parameters values take effect. For a cluster with 48 OSDs (Object Storage Daemons), redeploying Ceph may cost nearly half an hour.
(3) \textbf{Higher noise.}
Benchmarking results often contain stochastic noises that become much noticeable in distributed environments. 
%Noises can deviate from average performance by 150MB/s (2.5\%).

\begin{figure*}[!t]
	\centering
	\hspace{0.05\textwidth}
	\begin{subfigure}[b]{0.33\textwidth}
		\centering
		\includegraphics[width=\textwidth]{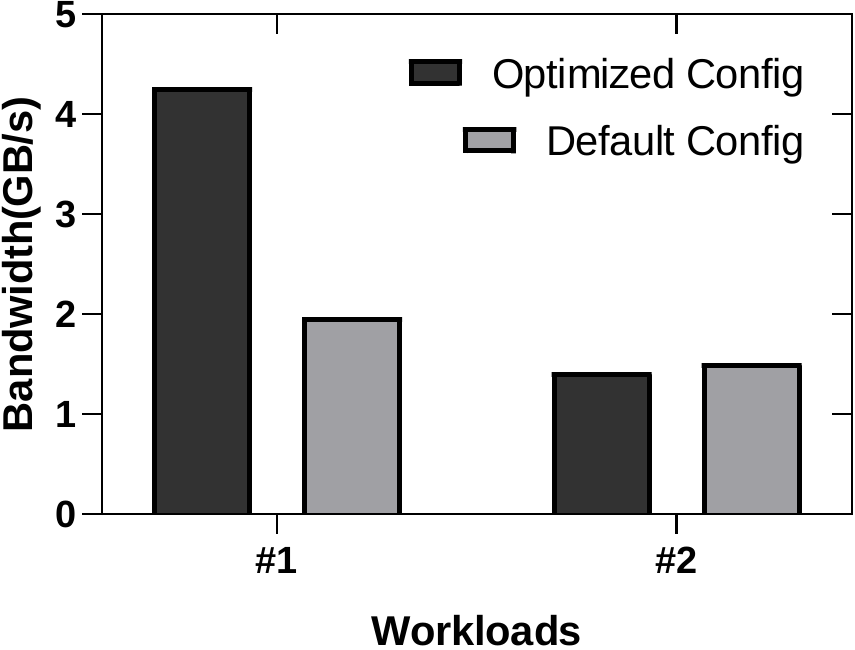}
		\caption{Non-reusable configuration}
		\label{fig:non-reusable}
	\end{subfigure}
	\hspace{0.15\textwidth}
	\begin{subfigure}[b]{0.35\textwidth}
		\centering
		\includegraphics[width=\textwidth]{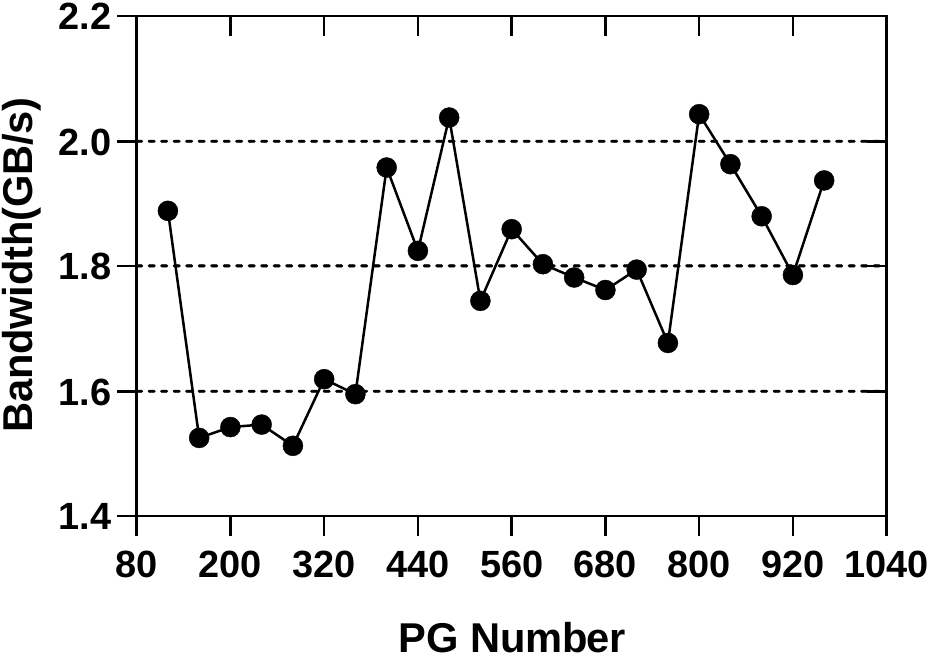}
		\caption{Non-linear performance effect}
		\label{fig:non-linear}
	\end{subfigure}
	\hspace{0.05\textwidth}
	% \vspace{-0.3cm}
	\caption{(a) shows performance measurements of the optimized configuration and the default configuration under two different. (b) shows how the bandwidth of Ceph changes when we alter the pg numbers in a pool.}
	% \vspace{-0.6cm}
\end{figure*}

In this paper, we propose \cn to automatically recommend near-optimal configurations.
Firstly, we provide general guidelines to solve the complex parameter constraints, generate a clean configurable parameter domain for later processes.
Secondly, we analyze different performance effects of Ceph parameters and find that only a small set of parameters have a significant impact on system performance.
Based on this observation, we rank parameter impacts and only care about the top knobs.
By tuning top parameters , we managed to significantly shrink the optimization time of \cn, while maintains the performance efficacy.
%To deal with the long evaluations time and the high evaluation noises, we analysis popular optimization techniques. 
Thirdly, we recommend using Bayesian optimization with Gaussian Process to effectively approximate system performance through noise-corrupted evaluation results.
Our evaluations show that \cn can significantly improve the average Ceph performance by 2.2$\times$ compared to the default configuration.

The rest of this paper is organized as follows. 
In section 2, we elucidate our motivation and difficulties in tuning a distributed storage system. 
In section 3, we elaborate on the design and implementation of our automatic tuning system. 
In section 4, we conduct evaluations and demonstrate that \cn can recommend configurations way better than the default and the expert. 
In section 5, we analyze the recent related works and explain the uniqueness and superiority of \cn. 
In section 6 and 7, we are looking forward to the future work and reach the conclusion.

\section{Motivation}
Ceph is a unified open-source distributed storage system that gains its popularity in the cloud environment~\cite{weil2006ceph}.
It comes with plenty of configurable parameters. 
But most users resort to using default settings, as the default configuration provided by developers is trusted to be "good enough". 
Tuning parameters is massively time-consuming and challenging for ordinary users, as profound experience and expertise in system internals are required. 
Even worse, Ceph lacks the functionality descriptions about parameters and the guidelines on how to tune thems, leaving users clueless.
The parameter information provided by the developer is too lacking, which also brings great difficulties to our research.

Unfortunately, default parameter settings are far from the optima and may result in poor I/O performance, especially for new hardware platforms, like NVMe SSDs. 
Default settings are mostly tuned for common commodity machines.
Users who pursue higher performance by using more powerful servers should reconfigure system settings to leverage the extra hardware resources.
Our evaluation shows that tuning parameters in Ceph can provide significant performance gains.
Thus we advocate that Ceph users should turn parameters based on their specific situations.

Traditionally, parameter tuning is done by system administrators.
They adjust settings, then measure system performance.
Based on the results, they tweak parameter values intuitively based on their experience and insight of the system.
But in distributed storage systems, like Ceph, tuning is much more challenging.
Also, the optimal settings are dependent on hardware and workload characteristics~\cite{cao2017variation}.
The best configuration for one workload may not perform as well for others.
In Figure~\ref{fig:non-reusable}, we optimize the configuration for workload $\#1$, then test the default and the optimized configurations on workload $\#2$.
Evaluations show that the optimized setting becomes ineffective on workload $\#2$.
The tuning complexity and non-reusable of optimal settings make manual tuning intractable.
Thus, to provide the best settings for various users, an efficient and effective automatic parameter tuning approach is needed in distributed storage systems.

Evaluation of distributed systems can be very time consuming, as restarting and redeploying clusters are needed to make new parameter values take effect.
For a Ceph cluster with 48 OSDs (Object Storage Daemons), one system evaluation may cost nearly half an hour, and 
the time increases as the cluster scales out.
Thus, we use a simulation-based approach to design a lightweight system for near-optimal configuration searching (Sec.~\ref{sec:overview}).
Evaluation processes are conducted in the small fixed-sized test environment. 
% Thus we can greatly reduce and limit the time required for one system evaluation.

During our study, we find new challenges in automatic parameter tuning of distributed storage systems.

(1) \textbf{Configuration constraints.} 
Complicated parameter value constraints exist inside the configuration space. 
Causing system performance changes not simply linear to the parameter value.
Figure~\ref{fig:non-linear} shows how bandwidth changes while we alter placement group numbers in a storage pool.
Such irregular and multi-peak correlations make it hard to achieve global optimal performance as we have to avoid those local optima.

Also, some constraints are neither documented by developers nor pinpointed by system log messages.
For example, in Ceph Luminous, the place group number is restricted from 30 to 250 per OSD.
This constrain is hardcoded in the system, but it is not displayed in the documentation provided by the developers.
Changing Parameter values must obey these constraints, as violating them can cause system failures or even crushes~\cite{Tianyin2013misconfig}.
However, in order to apply automatic algorithms, we need to provide a clean search domain that has explicitly defined boundaries and contains no misconfigurations.
To address this problem, we propose the parameter constraints solution to generate a well-defined parameter value domain under the constraints (Sec.~\ref{sec:constraints}).

(2) \textbf{Huge numbers of parameters.}
Distributed storage systems provide many more parameters, often exceed one thousand. 
Figure~\ref{fig:num_of_parameters} shows the growth of Ceph knobs over the last ten years.
With each release, more parameters are provided, while few are deprecated.
The newest Ceph Nautilus even comes with 1536 parameters.
Searching optimal settings in such enormous parameter space is still challenging for popular black-box optimization techniques.
This motivates us to solve this problem by ranking parameter impacts and search optimal configuration only with the top set of parameters (Sec.~\ref{sec:parameter_ranking}).

(3) \textbf{Higher noise.}
Benchmark results of storage systems often contain stochastic noises, which become much more noticeable in the distributed environments. 
Based on our experiments, benchmark noises can deviate from system average performance by 150MB/s (2.5\%).
The conventional approach to deal with noise is to take the average of multiple tests.
But such approaches are unbearably time-consuming.
To handle the high stochastic noises, we use Bayesian optimization with Gaussian Process to effectively search optimal configurations through such noise-corrupted observations (Sec.~\ref{sec:config_rec}).

The tuning efficiency and efficacy difficulties make tuning Ceph parameters a labored and complicated task.
Worse, the optimal settings cannot be used across different hardware and workload.
Manual configuring for every user case is intractable, due to the long tuning time and the difficulties to find the global optima.
Thus, to provide the best configurations for various users, an efficient and effective automatic parameter tuning approach is highly needed in Ceph distributed storage system.

\section{Design and Implementation}
\subsection{System Overview}
\label{sec:overview}

%Whether some parameters take effect may depend on others who control which module or functionality to use. 
\begin{table*}[]
	\makegapedcells
	\centering
	\begin{tabular}{p{0.5\textwidth}|p{0.45\textwidth}}
		\hline
		\textbf{Parameter Constraints} & \textbf{Ceph Examples (Luminous)} \\ \hhline{==}
		1. Some Parameters are unconfigurable. & \textit{fsid} and \textit{mon\_host} are fixed at the startup and should not be tuned. \\ \hline
		2. Some parameters have strict boundaries. & In Ceph Luminous, \textit{pg} number is restricted to $[30,250]$ per OSD. \\ \hline
		3.Some parameters determine whether others take effect, as they control which module or functionality to use. 
		& \textit{osd\_objectstore} determines the backend type for OSDs, which can be \textit{blueStore}, \textit{fileStore}, \textit{memstore} or \textit{kstore}. \\ \hline
		4. Some parameter values are interdependent, like they must have a fixed sum or one must lower than the other. & The sum of \textit{bluestore\_cache\_kv\_ratio} and \textit{bluestore\_cache\_meta\_ratio} must not exceed 1. \textit{ms\_async\_max\_op\_threads} constraints the max value of \textit{ms\_async\_op\_threads}\\ \hline
		%\textit{ms\_async\_op\_threads} value is constrained by \textit{ms\_async\_max\_op\_threads}.
	\end{tabular}
	\caption{Summary of value constraints inside the parameter space.}
	\label{tab:constraints}
\end{table*}

\begin{figure}[h!]
	\centering
	\includegraphics[width=\linewidth]{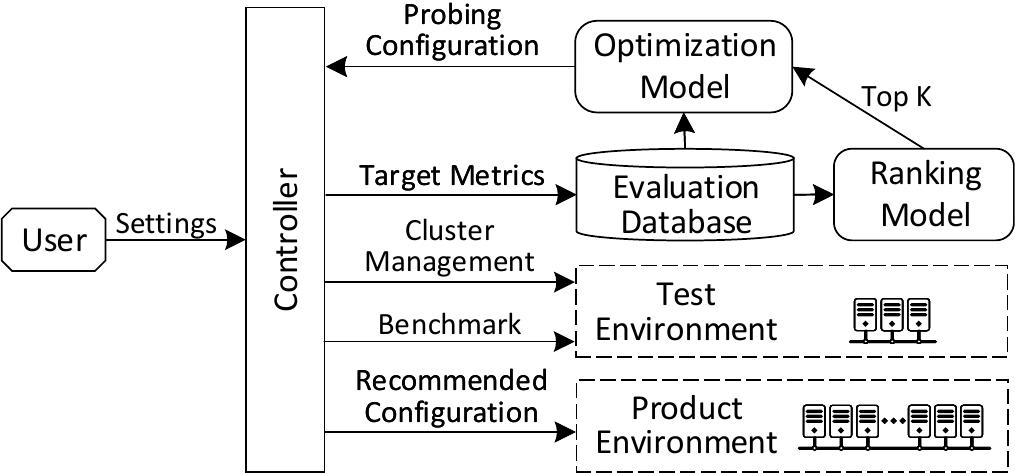}
	% \vspace{+0.4cm}
	\caption{\cn Framework}
	\label{fig:framework}
	% \vspace{+0.4cm}
\end{figure}

In this section, we present our automatic optimal configuration recommendation system, \cn, to overcome those parameters tuning challenges. Figure \ref{fig:framework} shows an overview of the workflow and the main components of the system.

\cn consists of two main parts, the controller and the machine learning (ML) models.
%customizing
The Controller accepts user settings like cluster setups, maximum iteration steps, and the number of top values used in ML models.
It manages the storage cluster, makes parameter changes take effect by injecting run-time commands, restart or re-deploy system services.
Controller also benchmarks the performance of the storage system and sends target metrics to the machine learning models.

Models consist of the ranking model and the optimization model.
All the system measurement results are stored in the evaluation database.
The ranking model processes all the evaluation results and produces a parameter ranking list according to their impact on system performance.
Based on the ranking, the optimization model uses the top K parameters to generates the search domain.
Then it probes different configurations and refines the learning model until the optimal settings are found, or the iteration reached the limit.

\cn adopts a simulation-based approach, which means it learns experiences and builds optimization models by evaluating the system performance of a small test cluster under the same simulated or initialized workload. 
Based on the trained models, it recommends optimal configurations for the large product storage cluster.
In this way, the process of finding optimal configurations would not interfere with the online service in the product environment. 
We find the test environment is much efficient to evaluate while preserving high accuracy in simulating dynamic behaviors os the large product cluster, as modern distributed storage systems provide good scalability.

\subsection{Parameter Constraints Resolution}
\label{sec:constraints}
We address the problem of constrained parameter value space.
%!!!Complicated Parameter value constraints exist inside the configuration space, which are ambiguously and not documented by system developers.
%Tuning parameters values must obey these constraints, as violating them may cause critical system failures or even crushes~\cite{Tianyin2013misconfig}.
%Preprocessing is an important stage 
In Table~\ref{tab:constraints}, we analyze and summarize the existing constraints inside the parameter domain.
We propose general preprocessing guidelines to deal with parameter constraints.
%for delimiting the parameters dependencies and simplifying the configuration space.
With this solver, we washout un-configurable parameters, prune unused ones, and set up the value boundaries.
Finally, we generate a clean and complete configurable parameter space, which contains no misconfigurations and has well-defined boundaries for later impact ranking and automatic tuning.

\textbf{Parameter washing:}
Unfortunately, parameter tuning lacks attention from Ceph developers.
The documents developers provided lack descriptions about parameters' functionalities and sadly did not contain all the knobs that could be tuned.
Thus, we analyze the configuration source code directly to get the complete parameters set.

Parameters that cannot be tuned like port numbers and IP addresses are mixed with configurable ones.
We statically analyze the variable names, data types, usage tags and descriptions of the parameters to address these problems.
We remove unconfigurable parameters like IP addresses, port numbers, path strings, and debugging used parameters.

\textbf{Parameter pruning:}
Modern distributed storage systems often consist of multiple layers of sub-modules and provide different implementations of the same functionalities for customization.
Different modules and implementations have unique parameters to control their behavior, and there are no shared parameters between them.
For example, Ceph provides \textit{Bluestore} and \textit{Filestore} as two different backends and uses \textit{osd\_objectstore} to determine which to use.
For one specific user case, some modules may not be used, and their parameters would not affect the system performance. Thus, we can ditch them to reduce the parameter space.

To resolve this constraint, we arrange module selecting parameters into an indexing structure and classify knobs based on which module and sub-module they belong to.
For the specific user case, we analyze modules it depends on, set values of module selecting parameters, and prune those unused ones to shrink the configuration space.
%only select parameters of those used components while

\textbf{Parameter boundary:}
Values of parameters are mixed with numerical and non-numerical ones.
We convert non-numerical values, such as \textit{boolean} or \textit{string}, into integers by mapping candidate values to 
consecutive indexes.

The other problem is that developers do not provide the boundary of most parameters.
But a limited search domain is necessary for optimization searching models.
Setting parameters boundaries intuitively based on their default values may be simple. 
But such a static approach cannot guarantee the optimal setting is included.
%In order to ensure the optimal configuration reachable, 
Thus, we dynamically enlarge the extent of parameters when the probing point comes near the boundary.

\begin{figure}[h]
	\centering
	\includegraphics[width=0.85\linewidth]{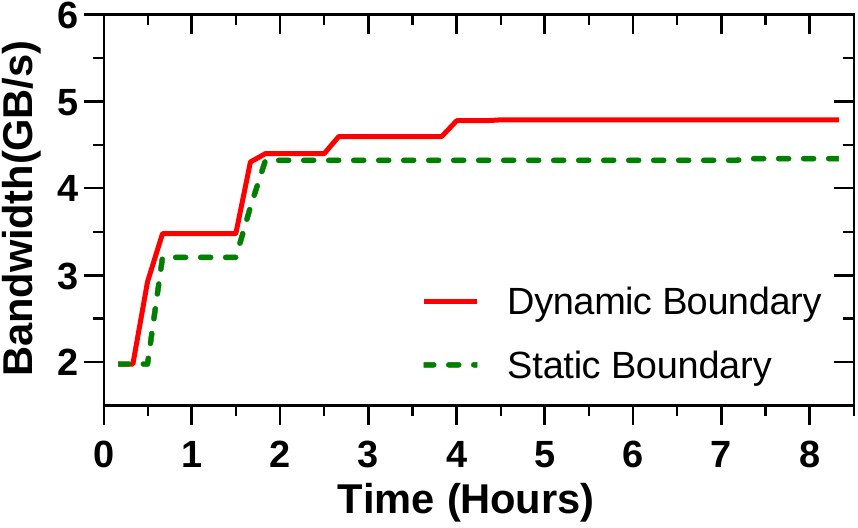}
	\caption{Dynamic boundary. The recommendation process of \cn with dynamic boundary and static boundary.}
	\label{fig:boundary}
\end{figure}

Figure~\ref{fig:boundary} shows that with the dynamic boundary strategy, our optimization algorithm can successfully find the global optimal results.
%Also parameters values must under certain

\begin{table*}[!t]
	\setcellgapes{2pt}%parameter for the spacing
	\makegapedcells
	\centering
	% \resizebox{\textwidth}{!} {
	% \begin{tabular}{l|l|c|c|c}
	\begin{tabular}{p{0.2\textwidth}|p{0.4\textwidth}|>{\centering\arraybackslash}p{0.06\textwidth}|>{\centering\arraybackslash}p{0.11\textwidth}|>{\centering\arraybackslash}p{0.08\textwidth}}
		\hline
		% \textbf{Configuration Parameters} & \textbf{Description} & \textbf{Value Type} & \textbf{Default Value} & \textbf{Value Range} \\ \hhline{=====}
		\textbf{Parameters} & \textbf{Description} & \textbf{Type} & \textbf{Default} & \textbf{Range} \\ \hhline{=====}
		\textit{\textbf{osd per nvme}} & The number of osds in a single NVMe SSD. & Integer & 1 & Dynamic \\ \hline
		\textit{\textbf{osd op num threads per shard ssd}} & The number of threads per shard for SSD   operations. & Integer & 2 & Dynamic \\ \hline
		\textit{\textbf{osd op num shards ssd}} & The number of shards for SSD operations. & Integer & 8 & Dynamic \\ \hline
		\textit{\textbf{bluestore cache size ssd}} & Default bluestore cache size for non-rotational (solid state) media & Integer & 3221225472 & Dynamic \\ \hline
		\textit{\textbf{bluefs alloc size}} & BlueFS instance of allocator is initialized with   bluefs alloc size & Integer & 1048576 & Dynamic \\ \hline
		\textit{\textbf{pg per osd}} & The place group number for each osd. & Integer & 100 & {[}30, 250{]} \\ \hline
		\textit{\textbf{objecter tick interval}} & None & Double & 5 & Dynamic \\ \hline
		\textit{\textbf{ms async rdma send buffers}} & How many work requestes for rdma send queue. & Integer & 1024 & Dynamic \\ \hline
		\textit{\textbf{osd max pgls}} & Maximum number of placement groups to list. & Integer & 1024 & Dynamic \\ \hline
		\textit{\textbf{osd loop before reset tphandle}} & Max number of   loop before we reset thread-pool's handle & Integer & 64 & Dynamic \\ \hline
		\textit{\textbf{osd op pq min cost}} & None & Integer & 65536 & Dynamic \\ \hline
		\textit{\textbf{osd max omap bytes per request}} & The max omap size for a single request. & Integer & 1073741824 & Dynamic \\ \hline
		\textit{\textbf{journaler write head interval}} & None & Integer & 15 & Dynamic \\ \hline
		\textit{\textbf{osd agent delay time}} & None & Double & 5 & Dynamic \\ \hline
		\textit{\textbf{osd agent max ops}} & Maximum number of simultaneous flushing ops per tiering agent in the high speed mode. & Integer & 4 & Dynamic \\ \hline
		\textit{\textbf{mgr mon bytes}} & None & Integer & 134217728 & Dynamic \\ \hline
	\end{tabular}
	% }
	\caption{Description of the top 16 configuration parameters generated by ranking.}
	\label{tab:top16}
\end{table*}

\subsection{Parameter Ranking}
\label{sec:parameter_ranking}

In this section, we provide parameter ranking to address the problem of the high-dimensional parameter space.
After the parameter preprocessing, we obtain a clean set of configurable parameters that may affect system performance afterward.
The challenge is that there are hundreds of configuration parameters in Ceph.
Searching for near-optimal settings inside such high dimensional configuration space is too challenging to be achieved.

During our experiments, we find that some parameters have a massive impact on system performance, while others seem to have no effect at all.
Based on this observation, we suggest only tuning the most influential parameters in the configuration recommendation process.
But another challenge is that we do not know what the most influential parameters are.
Ceph developers did not provide information about the influential parameters.
Due to the complexity of the Ceph system, we believe even developers have difficulty in determining the most influential parameters.
Doing experiments for each parameter to determine its influence on system performance is not feasible, as the number of parameters is huge.
Also, as we discussed before, system evaluation is very time-consuming in Ceph.
We have nearly thousands of parameters to analyze, yet we can only conduct a few tens to hundreds of evaluations.

In this paper, we propose to use machine learning techniques to quantify and rank parameter importance, and finally select the most important parameters.
We sample the entire parameter space randomly.
Then, we probe Ceph with these sample configurations and collect corresponding system performance.
Finally, based on the Lasso~\cite{zhang2008sparsity} method, we analyze the relationship between parameters and performance in the sample data and rank the parameter importance.
%we design to analyze performance effects of different Ceph parameters and ranking their impact. 

In machine learning, feature selection is the process of selecting a subset of the most relevant features.
Feature selection techniques are used to avoid the curse of dimensionality~\cite{highdimen}.
They are often used in domains where there are many features and comparatively few samples.
Feature selection methods are typically divided into filter method, wrapper method, and embedded method~\cite{introfeature}.
Filter methods are particularly effective in computation time and robust to overfitting~\cite{filter}.
However, they tend to select redundant features when correlation exists.
Wrapper methods can detect the possible interactions between features~\cite{wrapper}, but increase overfitting risk when the number of samples is insufficient.
They also need significant computation time when the number of features is large.
Embedded methods are proposed to combine the advantages of both previous methods, which perform feature selection as part of the model construction process.
% L1 and L2 regularization are introduced in the embedded feature selection method~\cite{l1l2}, which corresponding to Lasso and Ridge regression.
Lasso regression is a typical embedded feature selection method~\cite{l1l2}.

% We choose Lasso regression to rank parameter impacts.
% Choosing a suitable method to analyze the relations between parameters and performance is critical to the quality of ranking results.
% The major characteristic in our relation analysis problem is that the number of performance results that the algorithm can analyze is much fewer than the number of configurable parameters.

%The challenges to conduct such comes from two aspects:
%(1) Values of different parameters may obey some correlated constraints, resulting in multicollinearity~\cite{multicoll};
%Regression analysis is often used to estimate the relationships between one or more independent variables and a dependent variable.
%The Ordinary Least Squares (OLS) method is a standard approach in regression analysis by minimizing the sum of the squares of the residuals.
%is a standard approach in regression analysis, but 

Ridge regression and lasso regression are derived from the Ordinary Least Squares method, which is a standard approach in regression analysis by minimizing the sum of the squares of the residuals.
But the Ordinary Least Squares method may have a huge variance under such a situation, resulting in a biased inefficient model.
To solve this overfitting problem, Ridge regression uses the L2 penalty, which penalizes the sum of squared coefficients.
L2 penalty shrinks coefficients closer to zero to decreases the model complexity.
But it cannot zero out coefficients, thus ended up with all the features.
%and becomes capable to handle high dimensional data sets.
%not only minimize the sum of squared residuals but also (i.e. L2 penalty), .
%This behavior is called shrinkage.
%This way, Ridge regression decreases model complexity and works for high dimensional data sets.
%Lasso is conceptually similar to Rigid regression.
Lasso uses the L1 penalty which penalizes the sum of absolute values.
%For a high value of the regularization penalty, 
With L1 penalty, Lasso can zero many small coefficients out, thus exclude less relevant features and make the feature selection.
It makes Lasso work well in the high dimensional scenario.
% Instead, Rigid regression can not ditch useless parameters and cannot be used 
%Also, correlated variables are allowed in Lasso.
%During the model fitting, one of the correlated variables will gain a large coefficient, while the rest will be zeroed out.

Lasso has many advantages over other regularization and feature selection methods.
It is interpretable, stable, and computationally efficient~\cite{tibshirani1996regression,efron2004least}. 
There are many practical and theoretical researches backing its effectiveness as a consistent feature selection algorithm~\cite{tibshirani2018uniform,zhang2016materialization,tibshirani2016exact}.
Thus, we propose to use Lasso to quantify and rank parameter importance, and finally select the most important parameters.

We also need to preprocess the sample data before applying the lasso model.
Because, as far as we know, Lasso provides higher quality results when the features are continuous and have approximately the same order of magnitude.
\cn preprocesses the sample data in two steps.

First, \cn has to deal with the categorical parameters.
The categorical parameter is one that has two or more categories, but there is no intrinsic ordering to the categories.
For example, \textit{osd\_objectstore} has four values: \textit{blueStore}, \textit{fileStore}, \textit{memstore}, and \textit{kstore}.
And there is no agreed way to order these values from highest to lowest. 
In \cn, we transform categorical parameters into dummy variables.
For example, for a categorical parameter with $n$ values, \cn converts it into $n$ binary parameters that take on the values of zero or one.
Such converting may introduce more parameters.
But the number of categorical parameters are fairly small in ceph, only ten percent.
The performance degradation caused by converting can be ignored.

Second, \cn has to normalize the data values to have the same order of magnitude.
\cn uses log-transformation ($log1p$) to transform parameters and performance values.
After the log-transformation, the data values will have approximately the same order of magnitude.
Also, log transformation can decrease the variability of data and make data conform more closely to the normal distribution.
% $log1p$ (i.e. $ln(x+1)$) is a popular log-transformation approach.
% It returns the natural logarithmic value of $x + 1$.

%Such high-dimensional scenario makes  an ideal approach to 
%Based this fact and the properties of regression techniques above, we implement \textit{Lasso} to reduce useless parameters and rank effective ones according to its coefficient result.

\subsection{Configuration Recommendation}
\label{sec:config_rec}

In this section, we introduce the configuration recommendation process in \cn.
\cn leverages the experiment-driven black-box optimization techniques to search the near-optimal settings.
Black-box optimization suits well in our problem, as it views the complex system in terms of its inputs and outputs, and assumes obliviousness to the system internals.
Modeling-based techniques~\cite{liu2015mrcof}\TODO{} try to build efficacy and efficient performance prediction models via deep understanding and formalization of the system.
But they are hard to be implemented here, due to the complexity of distributed storage systems.

We model our problem as an optimization problem with the objective function: $m = F_{hw}(wl, conf)$.
The objective function is defined with following statements:
Given a Ceph system deployed on the certain hardware environment, $F_{hw}$; 
for a kind of workload, $wl$; 
suggest a configuration, $conf$, that optimize the target metric, $m$.
The target metric can be system bandwidth, latency or energy.

The experiment-driven black-box auto-tuning method contains two main units: \textit{Experiment Unit} and \textit{Search Unit}.
The \textit{Experiment Unit} takes parameter configurations as input.
It executes tests on the system with those configurations, then automatically collects results of target performance metrics.
The corresponding configurations and results are combined and provided to the \textit{Search Unit}.
Guided by an optimization algorithm, the \textit{Search Unit} selects the next configuration to try based on previously learned information.
This try and test procedure continues until it reaches the optimal results or the iteration limit.

%most suitable
Selecting the optimization algorithm is important to \cn's performance.
Combining the characteristics of Ceph, we analyzed and compared four widely studied optimization algorithms: Simulated Annealing (SA), Genetic Algorithms (GA), Reinforcement Learning (RL), and Bayesian Optimization (BO)~\cite{fred2011sa,behzad2013hdf5,zhang2019cdbtune,van2017dbml,SOPHIA}.
% Methods like Simulated Annealing (SA)~\cite{fred2011sa}, Genetic Algorithms (GA)~\cite{behzad2013hdf5}, Reinforcement Learning (RL)~\cite{zhang2019cdbtune} and Bayesian Optimization (BO)~\cite{van2017dbml,SOPHIA} are implemented to find near optimal configurations.
SA suggests the next configuration based on the states of the current system and does not learn from the old experience, which makes it unreliable to find the global optima.
GA generates a whole set of new configurations in each iteration.
%and requires system performance measurement for every one.
It requires much more system measurements than the others.
But system measurement in Ceph is very time-consuming, which makes GA less practical.
RL requires accurate data set to train the deep convolutional neural network.
When data noise exists, the learning process in RL can be extremely slow, because much effort is spent to unlearn the biased estimates~\cite{noiseRL}.

% GA and RL both leverage information from the past, but they all require plenty of experimental results.

% Selecting the most suitable optimization algorithm is an important yet difficult task.
%Bayesian optimization has been applied to varied designs, including database systems, streaming, storage systems and neural networks~\cite{}. 

%system performance evaluation results contain huge stochastic noises.
%Optimization algorithms who requires a precise data set (like Reinforcement Learning) cannot work well in such situations.
% To address the problem of highly noised evaluation results, we analyze and compare different black-box optimization methods. 
We observe that with Gaussian Process, Bayesian Optimization can approximate the objective function accurately through noise-corrupted observations.
Besides the tolerating of stochastic noises, BO utilizes the full history information of past evaluation results, which makes the optimal searching process more accurate.
Also, BO probes only one new configuration in each iteration.
It makes BO more efficient in searching.
Thus, we implement the optimization algorithm in \cn based on the Bayesian Optimization and Gaussian Process.
% to find the near optimal configurations in the distributed storage systems.
% as the kernel function

% making it particularly useful when evaluations are very costly.
%In distributed storage scenarios, the huge stochastic noises in  are the main difficulty we must take care of to select well-performed optimizing method. The first one is that evaluation results contain high stochastic noises. The other one is that evaluation times is very long, thus we need to reduce the evaluation numbers.

%There are four popular black-box optimization algorithms: Simulated Annealing (SA), Genetic Algorithms (GA), Reinforcement Learning (RL) and Bayesian Optimization (BO).
%Comparing to the others, 

%SA suggests next configuration based on the states of current system and does not learn from the old experience, which makes it unreliable to find the global optima.
%GA and RL both leverage information from the past. 
%GA generates a whole new set of configurations in each iterations, thus requires much more evaluation results.
%, and do the performance evaluation for every one,
%RL depends on huge and accurate data set to train the inner deep convolutional neural network, making it less practical.
%But in distributed storage systems, experiments are heavily time consuming and evaluation results are highly noised, which makes both GA and 

\section{Evaluation}
%In this section, we evaluate \cn's ability to automatically recommend the optimal configurations of Ceph distributed storage system.
In this section, we detail our evaluation of \cn. 
We first cover the experimental settings in Section 4.1. 
Section 4.2 provides an analysis of the top storage parameters generated by the parameter ranking process.
Section 4.3 demonstrates the efficiency of \cn.
We demonstrate \cn's recommended configuration outperforms the default and the expert ones in section 4.4.

\subsection{Experimental Setup}
We implement \cn on the Ceph \textit{RADOS}~\cite{weil2007rados} layer to improve the performance of the whole storage system from the very bottom.
\textit{RADOS} is the object storage layer in Ceph that provides the shared storage backend. 
% All user-consumable services like \textit{RGW}, \textit{RBD} and \textit{CephFS} use \textit{RADOS} to store data and metadate.
All user-consumable services like \textit{Ceph Object Store}, \textit{Ceph Block Device}, and \textit{Ceph File System} use \textit{RADOS} to store the data and metadata~\cite{cephdoc}.
\textit{RADOS} consists of two types of daemons: the Monitors and the OSDs.
The Monitors maintain the cluster map information.
The OSDs provide underlying object storage for all the user data.

\begin{table}[!h]
	% \setcellgapes{4pt}%parameter for the spacing
	\makegapedcells
	\centering
	\begin{tabular}{p{0.15\linewidth}||p{0.72\linewidth}}
	% \begin{tabular}{c||l}
		\hline
		Ceph & Luminous 12.2.12 \\ \hline
		OS & CentOS 7.4 with kernel 3.10 \\ \hline
		CPU & Intel Xeon Gold 6148 Processor \\ \hline
		Memory & 32GB Micron DDR4 DRAM $\times$ 8 \\ \hline
		Storage & 4TB NVMe SSD (P4510 Series) $\times$ 4 \\ \hline
		Network & ConnectX-5 100GbE (MT27800 Family) \\ \hline 
		Switch & Mellanox SN2700 \\
		\hline 
	\end{tabular}
	\caption{The hardware and software details. All the experiment machines have the same setup.}
	\label{tab:hardware}
\end{table}

We perform optimal configuration searching on the test environment and then test the recommended configuration on the product environment. 
The test environment is comprised of three hosts, one Monitor host, one OSD host and one Client host.
The product environment has six hosts, one Monitor host, three OSD host and two Client hosts.
Each host has four NVMe SSDs as storage devices.
Two clusters consist of machines with same hardware and software specification which is summarized in Table \ref{tab:hardware}.
%Each node consists of one Intel Xeon Gold 6148 Processor, 8$\times$32GB DDR4 DRAM, 4$\times$4TB NVMe SSD (P4510 Series) and one ConnectX-5 100GbE Ethernet Card.
%All servers run CentOS 7.4 with kernel 3.10 and Ceph Luminous 12.2.12.

%Ceph package ships with an  which exercises the cluster by way of librados, the low level native object storage API provided by Ceph, thus can be used to evaluate the performance of Ceph system at the Rados level. 

\begin{table}[!h]
	\makegapedcells
	\centering
	\begin{tabular}{p{0.05\linewidth}||>{\centering\arraybackslash}p{0.094\linewidth}>{\centering\arraybackslash}p{0.094\linewidth}>{\centering\arraybackslash}p{0.094\linewidth}>{\centering\arraybackslash}p{0.094\linewidth}>{\centering\arraybackslash}p{0.094\linewidth}>{\centering\arraybackslash}p{0.094\linewidth}}
	% \begin{tabular}{c||cccccc}
		\hline
		\textbf{ \# } & \textbf{Type} & \textbf{Prefill} & \textbf{Time} & \textbf{Size} & \textbf{Procs} & \textbf{Ops} \\ \hline
		1 & Rand & 240s & 120s & 16KB & 64 & 16 \\ \hline
		2 & Seq & 240s & 120s & 16KB & 64 & 16 \\ \hline
		3 & Write & - & 600s & 16KB & 64 & 16 \\ \hline
	\end{tabular}
	\caption{Radosbench workload settings. Rand and Seq refer to the radom read and the sequential read. For read benchmark we prefill objects at the start. Size stands for the Object size. Procs stands for the number of concurrent \textit{Rados bench} processes. Ops stands for concurrent I/O operations inside each process.}
	\label{tab:workload}
\end{table}

We use the Ceph inbuilt benchmark tool, the \textit{Rados bench}, to generate various workloads to evaluate the system performance.
There are three different workloads in our experiments.
Table \ref{tab:workload} shows the detailed settings of these workloads. 

For the simplicity of cluster management and the accuracy of performance evaluation, we disable Ceph authentication, debugging functions, and cache tiering agents for all experiments.
We set the replica to one and use a single storage pool for all \textit{Rados bench} clients.
Before the measurement of random read and sequential read performance, we prefill the pool with write operations. Prefill  takes more time than read as write speed is generally much slower.

\subsection{Parameter ranking}
\begin{figure*}[!t]
	\centering
	\includegraphics[width=\textwidth]{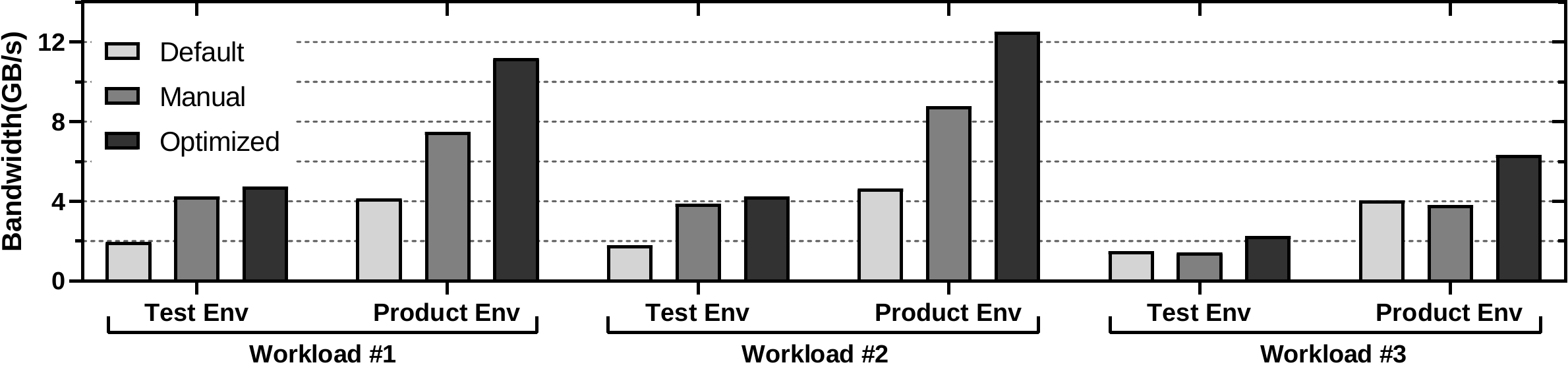}
	\caption{Performance measurement of Ceph under default, manual and \cn recommended configurations.}
	\label{fig:evaluation}
\end{figure*}
Here we calculate and analyze the importance of storage parameters generated by \cn's parameter ranking process.
We collect about three hundred system evaluations as the sample data.
All the evaluation results are under different configurations generated by random sampling.
After the collection and preprocessing, \cn calculates the importance of all the configurable storage parameters.
Table~\ref{tab:top16} lists the details about the top 16 most important parameters after the ranking.

\begin{figure}[!h]
	\centering
	\includegraphics[width=0.85\linewidth]{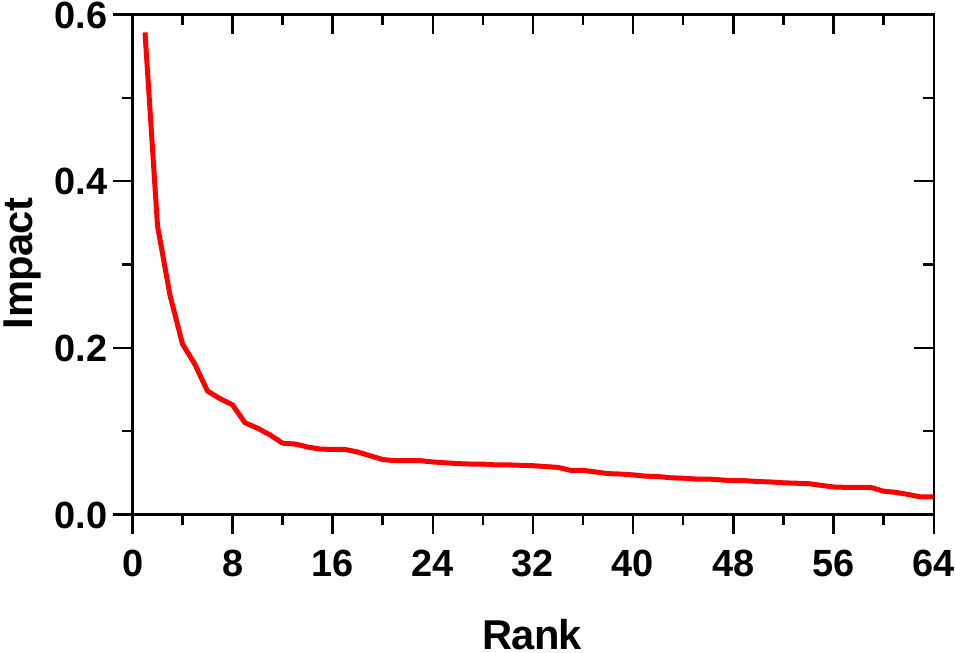}
	% \vspace{-0.3cm}
	\caption{Parameter ranking. By ranking, we quantify the parameter importance. This figure shows the result of the quantification.}
	\label{fig:coef_trend}
	% \vspace{-0.4cm}
\end{figure}

Figure~\ref{fig:coef_trend} presents the parameter ranking result by using \cn to analyze three hundred different evaluation results.
It is noted that the tendency line drops drastically, which informs us only the top set of parameters can significantly affect Ceph storage performance.
%we can find that parameter impact declines drastically, which demonstrates the performance of Ceph is actually affected by only a small subset of the top parameters.

\subsection{Efficiency of \cn}
%Time of Optimization Process
%Searching optimal settings in large parameter spaces may consume plenty of times.
Here we evaluate the time consumed by \cn to recommend near-optimal configurations.
In Sect.~\ref{sec:parameter_ranking},~\ref{sec:config_rec}, we rank the effect of Ceph parameters using Lasso and only use top parameters for later optimization.
To validate our design, we test \cn with top 64, 32, and 16 parameters under workload \#1.
We than measure the performance of recommended configurations and the consumed time during the optimization process. 

\begin{figure}[h]
	\centering
	\includegraphics[width=0.85\linewidth]{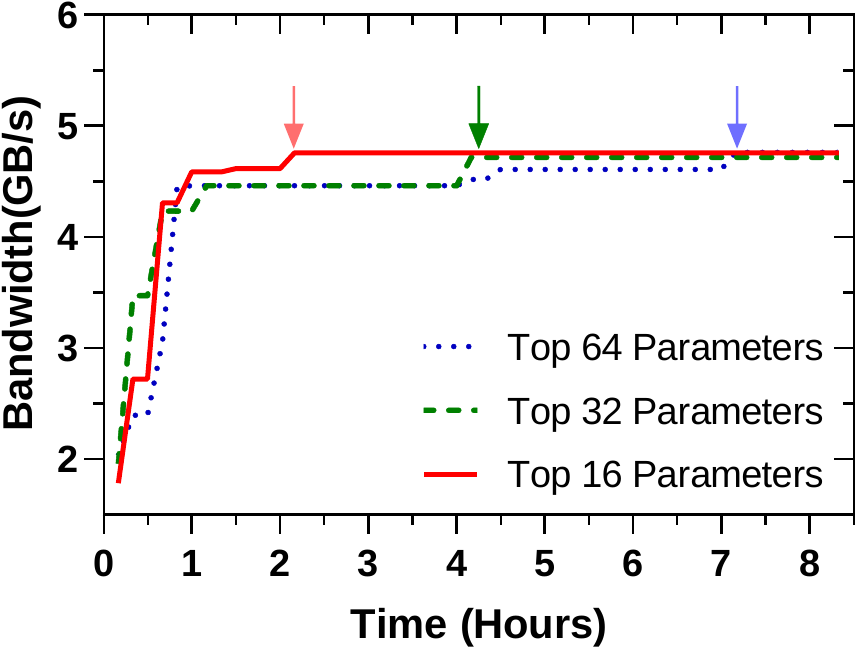}
	% \vspace{-0.2cm}
	\caption{The recommendation process of \cn with top 64, 32 and 16 parameters.}
	\label{fig:para_num_tune}
	% \vspace{-0.2cm}
\end{figure}

In Figure~\ref{fig:para_num_tune}, with the top 16 parameters, \cn uses only 2 hours to reach the optimal configuration.
While with the top 32 parameters, it takes nearly 7 hours to reach the optima.
Comparing to using the top 64 parameters, we find that using the top 16 ones only consumes 30\% of optimization time, while the performance of final recommendations has no apparent differences.
This is because the Ceph parameter impacts decrease drastically; most knobs have little effect on the performance.
Thus, by only using the top ones, we can significantly shrink the optimization time of \cn while maintains the performance efficacy.

\subsection{Effectiveness of \cn}

Here we evaluate the performance of the recommended configuration by \cn.
We generate the manual tuned configuration based on the Micron's storage solution for all NVMe Ceph~\cite{micron2018nvme}. 
Then, we compare the optimal configuration recommended by \cn with the default configuration and the manual configuration.

Figure~\ref{fig:evaluation} presents the measurement results for three different workloads under test and the product environments.
%the corresponding system evaluations.
%From the results we can find
%the optimized configurations are found through \cn in the test environment, then we evaluate them in the product environment.
Evaluation results show that the average Ceph performance is increased by 120\% with \cn compared to the default configuration.
%the recommended configurations by
And \cn can also out-perform manual tuned settings by 40\% averagely.

The default and the manual configurations are far from the optimal.
The manual setting could even impair system performance under specific workloads.
That is because configuration effects are highly related to the hardware and the workload characteristics. 
It is also noted that the recommended settings based on the test environment work similarly well in the large product environment, due to the excellent scalability of Ceph. 
%By comparing the test and the product environments, we demonstrate that 

\section{Related works}
In this section, we describe the related auto-tuning studies.
In recent years, several studies were made to automate the tuning of all kinds of computer systems~\cite{lu2019speedup}.
\TODO{list citation}
Jian \etal~\cite{tan2019ibtune} use neural networks to optimize the memory allocation of database instances, by adjusting buffer pool sizes dynamically according to the miss ratio.
Ashraf \etal~\cite{SOPHIA} perform a cost-benefit analysis to achieve long-horizon optimized performance for clustered NoSQL DBMS in the face of dynamic workload changes.
Ana \etal~\cite{Selecta} try to recommend near-optimal cloud VM and storage hardware configurations for target applications based on sparse training data.
Black-box optimization are used, as they view the system in terms of its inputs and outputs and assume obliviousness to the system internals.
Methods like Simulated Annealing~\cite{fred2011sa}, Genetic Algorithms~\cite{behzad2013hdf5}, Reinforcement Learning~\cite{zhang2019cdbtune}, and Bayesian Optimization~\cite{van2017dbml,SOPHIA} are implemented to find near-optimal configurations.

Zhen \etal~\cite{cao2018atc,cao2017variation,cao2019inter} tries to auto-tune storage systems to improve I/O performance.
They summarize the challenges in tuning storage parameters and then perform analysis of multiple black-box optimization techniques on the storage systems.
Their works mainly focus on the local storage systems, often with less than 10 parameters.
While in our work, we focus on the distributed storage systems and find new challenges, such as the configuration constraints, the huge numbers of parameters, and the higher noise.

Caver~\cite{Carver} also tries to solve the challenge of the large number of parameters and exponential number of possible configurations.
Like \cn, Caver proposes to focusing on a smaller number of more important parameters.
Inspired by CART~\cite{CART}, Carver uses a variance-based metric to quantify storage parameter importance.
Carver is designed for categorical parameters, as they find most parameters in local storage systems are discrete or categorical.
But we observe the exact opposite in Ceph, as most configurable parameters are continuous (about 90 percent).
From Table~\ref{tab:top16} we can find that all the top 16 parameters are continuous.
Although there are discretization techniques that can break continuous parameters into discrete sections, feature-selection results depend heavily on the quality of discretization~\cite{FeSelection}.
Thus, Carver is not suitable for our problem.
Different from Carver, \cn leverages Lasso to choose important knobs.
Lasso can provide higher quality results for continuous parameters.
And a small number of categorical parameters would not degrade the parameter ranking's performance.
% are very few, Lasso can works pretty well by converted categorical parameters into binary dummy features.

% and DAC~\cite{DAC} both
SmartConf~\cite{SmartConf} try to auto-adjust performance-sensitive parameters in the distributed in-memory computing system, spark, to provide better performance.
SmartConf uses a control-theoretic framework to automatically set and dynamically adjust parameters to meet required operating constraints while optimizing other system performance metrics.
% Althrough SmartConf can improve system performance without the out-of-memory and the out-of-disk problem, it also has great limitations.
But SmartConf does not work if the relationship between performance and parameter is not monotonic.
While in our cases, based on Figure~\ref{fig:non-linear}, those relationships can be irregular and multi-peak.
Unlike SmartConf, \cn uses machine learning techniques, which is a better fit for such complicated configuration space to find near-optimal settings.

DAC~\cite{DAC} finds that the number of performance-sensitive parameters in spark is much larger than previous related studies (more than 40 vs. around 10).
% simple sub-models
DAC combines multiple individual regression trees in a hierarchical manner to address the high dimensionality problem.
To reduce the modeling error, the hierarchical modeling process requires a large number of training examples, which is proportional to the number of parameters.
But there are hundreds of performance-related parameters in our problem comparing to 40 in DAC.
Modeling such a high-dimensional system with DAC would require hundreds of hours to collect training examples, which is impractical.

%in \cn, we addresss the hundreds of configurable parameters which is 10 times bigger

%\cn is different from this work since 

%Another class of relatedworks attempt to optimize the local file systems.
% \input{future.tex}
\section{Conclusion}
%In modern distributed storage systems, default configurations provided by developers are often sub-optimal for specific user cases.
%Turning parameters can provide significant performance gains, but comes with tough challenges.
Configuration constraints and huge numbers of parameters are difficult challenges in automatic configuration recommendation in distributed storage systems.
We provide general guidelines to resolve such constraints and using ranking strategy to get top influential parameters.
Our simulation-based approach not only leaves online services undisturbed, but also produces high-quality configurations.
Evaluations show that recommended configurations by \cn perform much better than default and the manual configurations.

%During our study,
%In this work, we provide general guidelines to preprocess ambiguous parameter spaces.

%We implement this automatic configuration recommendation framework for Ceph distributed storage system. 
%\cn preprocesses the parameters based on the guidelines to generate complete tunable search domain.
%Then, it analysis parameters impact based on the past observations and generate a ranking list. 
%Finally, it uses the top parameters and BO to automatically recommend the optimal configuration. 
%Evaluations show that recommended configurations by \cn can produce significant 2x performance improvement. 
%Evaluations show that the average Ceph performance is increased by 120\% with \cn compared to the default configuration.

\bibliographystyle{abbrv}
\bibliography{paper}

\end{document}